%% file: main_L.tex
\documentclass[aps,prd,twocolumn,superscriptaddress,tightenlines,nofootinbib]{revtex4-1}
\usepackage[T1]{fontenc}
\usepackage{microtype}
\usepackage[textsize=scriptsize,backgroundcolor=red!70,linecolor=red]{todonotes}
\usepackage{booktabs}
\usepackage{amsmath}
\usepackage{amsfonts}
\usepackage{amssymb}
\usepackage{hyperref}
\usepackage[all]{hypcap}
\usepackage{paralist}
\usepackage{multirow}
\usepackage{graphicx}
\usepackage{dcolumn}
\usepackage{bm}
\usepackage{epsf}
\usepackage{comment}
\usepackage{ulem}

\newcommand*{\rom}[1]{\expandafter\@slowromancap\romannumeral #1@}

\def\beq{\begin{equation}}
\def\eeq{\end{equation}}
\def\beqa{\begin{eqnarray}}
\def\eeqa{\end{eqnarray}}

\newcommand{\pc}[1]{\textcolor{orange}{[PC: #1]}}

\begin{document}
\DeclareGraphicsExtensions{.eps,.ps}

\title{Bounds on nonlinear effective field theories via resurgent relative entropy}
	
\author{Pietro Conzinu,}
\email[Email Address: ]{p.conzinu@ssmeridionale.it}
\affiliation{Scuola Superiore Meridionale, Via Mezzocannone 4, 80138 Napoli, Italy}
\affiliation{INFN-Napoli, Complesso Universitario di Monte S. Angelo, Via Cinthia Edificio 6, 80126 Napoli, Italy}
	
\author{Daiki Ueda}
\email[Email Address: ]{daiki.ueda@campus.technion.ac.il}
\affiliation{Physics Department, Technion -- Israel Institute of Technology, Haifa 3200003, Israel}
	
\begin{abstract}
We study a class of nonlinear effective field theories (EFTs) in which relative entropy organizes an infinite tower of higher-dimensional operators with factorially growing coefficients.
Using its Borel-resummed expression, we derive bounds on EFT coefficients: under the factorial-growth ansatz, the non-negativity of relative entropy in the weak-background regime constrains the sign of their assumed asymptotic growth, while a negative real part beyond this regime signals perturbative limitations and possible sensitivity to nonperturbative effects.
In fermionic QED, analytic continuation from Euclidean to Minkowski spacetime yields a controlled one-loop example: such negativity accompanies the Schwinger instability.
\end{abstract}
	
\maketitle

\noindent{\bf Introduction.} --- Effective field theories (EFTs) provide a unifying framework for low-energy quantum field theory, systematically organizing the effects of heavy degrees of freedom~\cite{Georgi:1993mps,Burgess:2007pt}.
When the effective description becomes nonlinear, characterized by a formal infinite series in the EFT expansion, assessing its validity and internal consistency~\cite{Pham:1985cr,Adams:2006sv} becomes increasingly subtle, especially beyond conventional perturbative regimes.
Establishing robust and model-independent bounds on nonlinear EFTs is therefore a fundamental challenge, calling for theoretical tools capable of probing nonperturbative structures.

In recent years, it has been shown that the non-negativity of the relative entropy~\cite{Kullback:1951zyt,10.2996/kmj/1138844604,RevModPhys.50.221}, evaluated within the EFT expansion to leading order, leads to nontrivial constraints on certain classes of EFTs~\cite{Cao:2022iqh,Cao:2022ajt,Ueda:2024cyf}\footnote{In Ref.~\cite{Fernandez-Sarmiento:2025tpy}, related directions were explored.}.
In particular, with appropriately chosen probability distributions, this non-negativity can be interpreted as reflecting the unitarity of the underlying theory.
Standard amplitude-based positivity bounds constrain Wilson coefficients through dispersion relations based on analyticity, crossing, and unitarity, together with locality, causality, and suitable high-energy behavior~\cite{Adams:2006sv,deRham:2017avq,Tolley:2020gtv}.
Our derivation instead uses relative-entropy non-negativity without an amplitude dispersion relation, suggesting that the two approaches should be viewed as complementary. 
A concrete connection between the two approaches can be seen when dimension-eight terms dominate on the chosen weak backgrounds.
In this regime, previous relative-entropy analyses have derived constraints on the Wilson-coefficient combinations associated with scalar $(\partial\phi)^4$ and gauge $F^4$ operators that are consistent with standard forward-limit positivity bounds.

Beyond this leading-order overlap, relative entropy organizes the background-projected nonlinear tower; assuming the factorial-growth ansatz introduced below, Borel--Laplace resummation and weak-background non-negativity constrain the sign of its assumed asymptotic coefficient growth.
As an application, we study fermionic QED to test the bounds derived here.
A detailed analysis of fermionic QED, along with extensions to scalar QED, the DBI model, and power-growing EFTs beyond the factorial-growth scenario considered here, is presented elsewhere~\cite{Conzinu:2026qgr}.
\vspace{0.3\baselineskip}

\noindent{\bf Formal setup.} --- Throughout this Letter, we focus on nonlinear EFTs satisfying the following assumptions: (i) the nonlinear EFT corrections arise from interactions between UV and IR degrees of freedom in a theory described by a self-adjoint Hamiltonian; (ii) symmetry and suitable backgrounds reduce the relative entropy, after canonical normalization, to a background-projected tower of higher-dimensional operators, as shown explicitly for shift-symmetric scalar and pure-gauge EFTs; and (iii) for the explicit bounds derived here, we assume the factorial-growth ansatz specified below and require contributions not captured by the local weak-background expansion to remain parametrically subleading in the weak-background limit. 
Assumption~(iii) is not required for the relative-entropy construction itself, but is invoked only to derive the explicit constraints on the asymptotic coefficient growth presented below. 
Power-growing EFTs are analyzed in Ref.~\cite{Conzinu:2026qgr}.

We first rely solely on assumption (i) and study the non-negativity of the relative entropy between the interacting UV theory (the target theory) and a corresponding heavy--light-decoupled reference theory, introduced to isolate the EFT effects.
Following Refs.~\cite{Cao:2022iqh,Cao:2022ajt,Ueda:2024cyf}, we define $
S(\rho_{\rm R}\|\rho_{\rm T}) \equiv {\rm Tr}\,\big[\rho_{\rm R}\ln \rho_{\rm R}-\rho_{\rm R}\ln \rho_{\rm T}\big] \ge 0$,
where $\rho_{\rm T}$ and $\rho_{\rm R}$ are the density operators of the interacting target theory and the heavy--light-decoupled reference theory, respectively.  
The non-negativity holds for positive semi-definite, normalized density operators, and is saturated if and only if $\rho_{\rm R} = \rho_{\rm T}$, quantifying the information that distinguishes the interacting theory from its reference counterpart and isolating the EFT effects originating from the UV interactions.

Specifically, we consider a regular family of UV theories described by self-adjoint Hamiltonians $H_g=H_0+V_g$, where $V_g=\sum_{n=1}^{\infty}g^n v^{(n)}$ encodes the heavy--light interaction.
Here, $H_0$ is an auxiliary reference theory rather than a separately realized physical UV theory: it contains the same heavy and light degrees of freedom as $H_g$, retaining interactions within each sector while removing the physical heavy--light interaction.
We require $H_g\to H_0$ regularly as $g\to0$.
Thus, varying $g$ specifies a formal decoupling path in theory space and need not correspond to an experimentally tunable coupling.
EFTs for which this decomposition or regular decoupling path cannot be consistently defined lie outside the present framework.
The nonlinear EFT terms encode the cumulative effects of $V_g$ generated upon integrating out the heavy sector.
Following Refs.~\cite{Cao:2022ajt,Cao:2022iqh,Ueda:2024cyf}, we define the density operators as $\rho_{\rm R} = e^{-\beta H_0}/Z_0$ and $\rho_{\rm T} = e^{-\beta H_g}/Z_g$, with partition functions $Z_0 = {\rm Tr}\,[e^{-\beta H_0}]$ and $Z_g = {\rm Tr}\,[e^{-\beta H_g}]$, where $\beta$ is the inverse temperature if interpreted canonically.  
For the prescribed Minkowski background, the relative-entropy inequality applies whenever $H_0$ and $H_g$ are self-adjoint and $e^{-\beta H_0}$ and $e^{-\beta H_g}$ have finite traces, so that they define positive, normalized density operators. In this domain, $S(\rho_{\rm R}\Vert\rho_{\rm T})\geq0$.
The effective-action expression obtained by analytic continuation to an unstable background, such as a constant electric field, need not retain this density-operator representation.

The relative entropy between $\rho_{\rm R}$ and $\rho_{\rm T}$, which encodes the information transferred from the UV theory to the EFT, can be written as
\begin{align}
    S(\rho_{\rm R}\|\rho_{\rm T}) = W_0 - W_g + {\rm Tr}\,[\rho_{\rm R}\, \beta\, V_g] \ge 0\,,
    \label{eq:quanrel}
\end{align}
where $W_0 \equiv -\ln Z_0$ and $W_g \equiv -\ln Z_g$ are the effective actions.
The regular decoupling limit implies $\lim_{g\to 0} \rho_{\rm T} = \rho_{\rm R}$.  
Differentiability along this path then yields $({d W_g}/{dg})_{g=0} = {\rm Tr}\,[\rho_{\rm R}\, \beta\, v^{(1)}]$.
For completeness, a derivation of Eq.~\eqref{eq:quanrel} is provided in Sec.~\ref{supp:1} of the Supplemental Material; see also Refs.~\cite{Cao:2022iqh,Cao:2022ajt,Ueda:2024cyf}.
Throughout this Letter, we work in the limit $\beta \to \infty$, establishing the connection between the relative entropy and the zero-temperature effective action.
We focus on the infrared regime where heavy degrees of freedom are integrated out, so that $W_g$ represents the EFT of the light sector.
Within the restricted class specified by assumptions~(i) and (ii), the effective action for background light fields can be written, after canonical normalization, as $W_g=W_0-W_{\rm nonlin}$, where $W_{\rm nonlin}\to0$ as $g\to0$. 
It encodes the cumulative connected effects induced by $V_g$.
An individual Wilson coefficient may receive multiple-insertion and multiloop contributions. 
For EFTs considered here, as shown explicitly in the Supplemental Material for the shift-symmetric scalar and pure-gauge EFTs, the symmetry assumptions and chosen backgrounds restrict the $V_g$-induced renormalizable response to kinetic renormalization, which canonical normalization removes.
Hence the third term in Eq.~\eqref{eq:quanrel} does not contribute, and assumption~(ii) gives $S(\rho_{\rm R}\Vert\rho_{\rm T})=W_{\rm nonlin}$ (see Secs.~\ref{supp:2} and \ref{supp:3} of the Supplemental Material).

The Hamiltonians $H_0$ and $H_g$ refer to the prescribed Minkowski backgrounds, whereas the corresponding effective-action functions are evaluated through an auxiliary Euclidean path integral and then analytically continued in the background fields.
When the auxiliary Euclidean construction admits a representation in terms of positive, normalized density operators, the corresponding relative entropy is non-negative.
After continuation to an unstable background, such as a constant electric field, vacuum decay can obstruct the stationary density-operator representation required for the relative-entropy inequality, and the continued effective-action expression may become complex.
Such a continued quantity need not itself represent a relative entropy between positive, normalized density operators.
Nevertheless, in what follows, we denote this formal analytic continuation by the same symbol $S(\rho_{\rm R}\Vert\rho_{\rm T})$ and study the sign of its principal-value real part as a diagnostic of sensitivity to nonperturbative effects, whose physical interpretation is model-dependent and may include instabilities.
\vspace{0.3\baselineskip}

\begin{figure}[t!]\includegraphics[width=\linewidth]{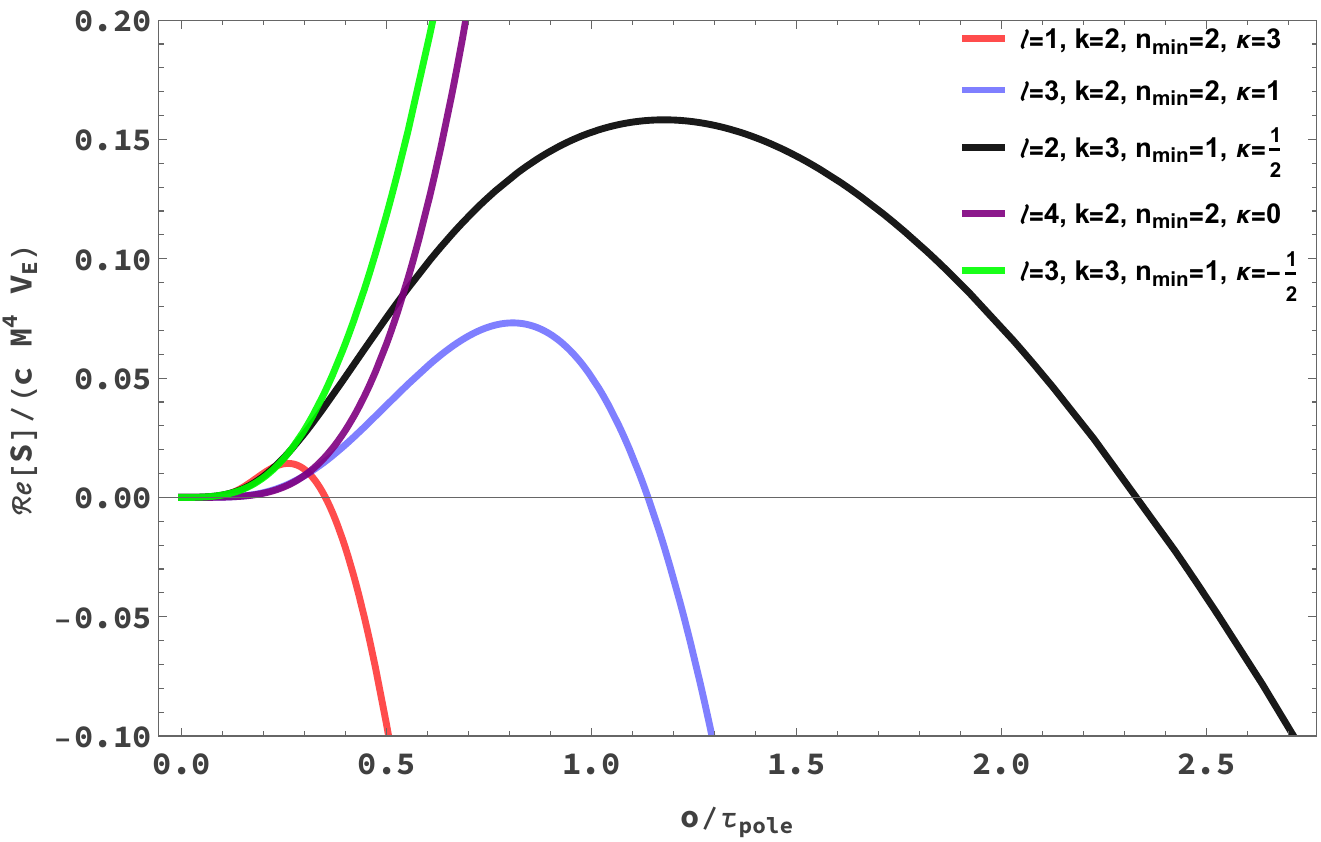}
\caption{
Real part of the analytically continued Borel-resummed expression in Eq.~\eqref{eq:resu_rel}, $\operatorname{Re}[S]/(CM^4 V_{\rm E})\equiv \operatorname{Re}S(\rho_{\rm R}\Vert\rho_{\rm T})/(CM^4V_{\rm E})$ with $V_{\rm E}\equiv \int d^4x_{\rm E}$, as a function of $o/\tau_{\rm pole}$ for $(o/\tau_{\rm pole})^k>0$, shown for several values of $\ell$, $k$, $n_{\min}$, and $\kappa=k(n_{\min}-1/2)-\ell+1$.
In the weak-background regime, $\operatorname{Re}[S]/(CM^4V_{\rm E})$ is positive, whereas in the strong-background regime its sign is controlled by $\kappa$.
For a generic EFT, a negative real part in the latter regime indicates a limitation of the resummed perturbative sector and may signal sensitivity to nonperturbative physics, including a possible physical instability.
}
\label{fig:Skln}
\end{figure}

\noindent{\bf Nonlinear EFTs with factorial growth.} --- We now adopt the IR perspective and consider nonlinear EFTs satisfying the factorial-growth ansatz in assumption~(iii).
This behavior reflects underlying nonperturbative physics and enables Borel--Laplace resummation of the relative entropy.
Under assumptions (i) and (ii), the relative entropy is
\begin{align}
    S(\rho_{\rm R}\|\rho_{\rm T})=W_{\rm nonlin}=\int d^4x_{\rm E}\, \mathcal{L}_{\rm nonlin}(o^k)\,,\label{eq:qed_rel}
\end{align}
where $\mathcal{L}_{\rm nonlin}(o^k)$ has the formal weak-background EFT expansion $\mathcal{L}_{\rm nonlin}(o^k)=M^4\,\sum_{n=n_{\rm min}}^\infty c_n\,o^{kn}$,
with $o^k$ a dimensionless normalized operator constructed from the background light fields,
$k,n_{\rm min}\ge1$ integers, and $M$ a characteristic UV mass scale.
In Minkowski spacetime, the EFT Lagrangian in a background is given by
$\mathcal{L}_{\rm EFT} = \mathcal{L}_0 + \mathcal{L}_{\rm nonlin}(o^k)$,
where $\mathcal{L}_0$ contains the standard renormalizable operators.
Within the class considered here, canonical normalization together with constant-gradient or Abelian/color-aligned constant-field-strength backgrounds reduces the background-projected $V_g$-induced nonlinear sector to an expansion in a single constant operator $o^k$.
Potential, mass, and other relevant terms, as well as $g$-independent light-sector operators, lie outside this identification.
Operators containing additional derivatives, and Lorentz or color structures that vanish or become degenerate on the chosen backgrounds, are not separately constrained.
Thus, Eq.~\eqref{eq:qed_rel} is a background-projected relation within this restricted sector, and extensions using other reference theories or backgrounds remain open (see Sec.~\ref{supp:3}).
Note that $W_g$ is defined from the Euclidean effective action
via analytic continuation of the background fields,
so that $\mathcal{L}_{\rm nonlin}$ enters the Minkowski-space Lagrangian with a plus sign\footnote{Since $W_g$ is obtained by analytically continuing the Euclidean effective action, with all background fields defined in Minkowski spacetime, only the overall sign reflects the Euclidean signature.}.
Under assumption~(iii), by the factorial-growth ansatz, we mean $c_n=C(\tau_{\rm pole})^{-kn}(kn-\ell)!$ for all $n\geq n_{\min}$, with $C\in\mathbb{R}$, $\ell\in \mathbb{Z}$, and $kn_{\rm min}-\ell\geq 0$.
This factorial-growth ansatz gives the formal weak-background series zero radius of convergence and allows its Borel transform and Borel--Laplace resummation to be evaluated in closed form.
If the relation holds only asymptotically, these expressions instead describe the leading large-order sector of the perturbative tower.
Defining the Borel transform as $ \mathcal{B}(\tau)=\sum_{n=n_{\rm min}}^\infty \tfrac{c_n }{(kn-\ell)!}\,\tau^{kn-\ell}$ (see, {\it e.g.}, Sec.~\ref{supp:Borel} of the Supplemental Material), the singular part can be written as  $\tau^\ell\, \mathcal{B}_{\rm sing}(\tau)
    =C\, (1-(\tau/\tau_{\rm pole})^k)^{-1}$. 
We employ the Borel--Laplace resummation, with $\tau \propto o$ (see Eq.~\eqref{eq:resu_rel}), and the absence of terms with $n<n_{\rm min}$ in the nonlinear EFT expansion fixes the corresponding coefficients of the Borel transform to vanish.
Within the factorial-growth ansatz, the Borel transform is given by\footnote{This Borel transformation is uniquely fixed under the assumption that the higher-order terms beyond $o^{k n_{\rm min}}$ in the EFT expansion are controlled solely by the coefficients $c_n=C\, (\tau_{\rm pole})^{-kn}\,(kn-\ell)!$.
If $c_n = C\, (\tau_{\rm pole})^{-kn}\,(kn-\ell)!$ captures only the asymptotic behavior, the resulting Borel transform is not unique and provides an approximate description of the leading large-order sector.
}
\begin{align}
    \tau^\ell\, \mathcal{B}(\tau)
    =
    C\, \frac{(\tau/\tau_{\rm pole})^{k n_{\rm min}}}{1-(\tau/\tau_{\rm pole})^k}\,,\label{eq:tauB}
\end{align}
which ensures the vanishing of the first $n_{\rm min}$ coefficients in the expansion around $\tau=0$.
The corresponding Borel-resummed expression is
\begin{widetext}
\begin{align}
    S(\rho_{\rm R}\|\rho_{\rm T})
    =
    \int d^4x_{\rm E}\, \mathcal{L}_{\rm nonlin}(o^k)= C\,M^4 \int d^4x_{\rm E} 
    \begin{cases}
    \int_0^\infty e^{-t} t^{-\ell}\, \frac{ (o t/\tau_{\rm pole})^{k n_{\rm min}}}{1-(ot/\tau_{\rm pole})^k}dt\,, & (o/\tau_{\rm pole})^k<0\,,
    \\
    \mathcal{P}\, \int_0^\infty e^{-t} t^{-\ell}\, \frac{ (o t/\tau_{\rm pole})^{k n_{\rm min}}}{1-(ot/\tau_{\rm pole})^k}dt \pm i\frac{\pi}{k}\, \frac{e^{-|\tau_{\rm pole}/o|}}{|\tau_{\rm pole}/o|^{\ell-1}}\,, &  (o/\tau_{\rm pole})^k>0\,,
    \end{cases}\label{eq:resu_rel}
\end{align}
\end{widetext}
where we have employed the Borel--Laplace resummation, $\mathcal{L}_{\rm nonlin}(o^k)=M^4\int_0^\infty e^{-t}o^\ell\mathcal{B}(ot)\,dt$.
For $(o/\tau_{\rm pole})^k<0$, the Borel integral does not encounter the nearest singularity along the integration contour.
Therefore, under the weak-background suppression condition in assumption~(iii), relative-entropy non-negativity for sufficiently small nonzero $o$ implies
\begin{align}
S(\rho_{\rm R}\|\rho_{\rm T})>0
\quad \Rightarrow \quad C\,(-1)^{n_{\rm min}}>0 \,.\label{eq:pos_mag}
\end{align}
Thus, under the factorial-growth ansatz, Eq.~\eqref{eq:pos_mag} constrains the sign of the assumed asymptotic coefficient growth.

By contrast, for $(o/\tau_{\rm pole})^k>0$, the Borel--Laplace integral is defined as a principal-value integral, $\mathcal{P}\,\int_0^\infty dt\,(\cdots)$, capturing the real part of the contribution from the nearest Borel singularity.
The resulting imaginary contribution exhibits a sign ambiguity (the Stokes ambiguity) due to the choice of integration contour, depending on whether the pole is avoided from above or below.
It therefore reflects the contour dependence of the resummation and the associated Stokes discontinuity.
Since perturbation theory alone does not, in general, fix this choice, an additional prescription is required to remove the resulting nonperturbative ambiguity.
If an imaginary part remains after the resummation prescription is fixed, the analytically continued quantity is complex and therefore can no longer itself be interpreted as a relative entropy between positive, normalized density operators.
In QED, this imaginary part encodes Schwinger pair production, whose vacuum instability prevents the analytically continued constant-field effective-action expression from admitting an interpretation in terms of positive density operators, as required for the relative-entropy inequality.
A negative principal-value real part may accompany such an instability, but it pertains to the analytically continued expression and does not constitute a violation of relative-entropy non-negativity.

We therefore analyze the real part of this analytically continued resummed expression for $(o/\tau_{\rm pole})^k>0$ in Eq.~\eqref{eq:resu_rel}.
While the imaginary part suggests a possible nonperturbative instability discussed above, the principal-value real part represents the resummed perturbative sector in $V_g$, whose finite-order contributions are real.
As shown analytically in Sec.~~\ref{sec:app:pos} of the Supplemental Material, with $\kappa\equiv k(n_{\min}-1/2)-\ell+1$, the real part of the Borel integral remains positive for $\kappa\neq0$ in the weak-background regime, $|o|\lesssim|\tau_{\rm pole}/\kappa|$.
For $\kappa=0$, it remains non-negative throughout the domain considered.
For $\kappa>0$, by contrast, it can become negative in the strong-background regime, $|\tau_{\rm pole}|\ll|o|$, where the contribution from the nearest Borel singularity is no longer negligible.
In Fig.~\ref{fig:Skln}, the numerical results for the real part of the analytically continued resummed expression, with $(o/\tau_{\rm pole})^k>0$, are shown as functions of $o/\tau_{\rm pole}$ for different values of $\ell$, $k$, $n_{\rm min}$, and $\kappa$.
Therefore, for $(o/\tau_{\rm pole})^k > 0$ in the perturbative regime, imposing relative-entropy non-negativity as a perturbative consistency condition gives
\begin{align}
    {\rm Re}\, S(\rho_{\rm R}\|\rho_{\rm T})>0 
    \quad \Rightarrow \quad 
    C>0\,.\label{eq:pos_ele}
\end{align}
Equation~\eqref{eq:pos_ele} is a perturbative weak-background sign constraint on the assumed asymptotic growth.
As $o\to0$, the Borel--Laplace-resummed expression reproduces the formal weak-background series asymptotically, so its leading behavior agrees with the perturbative result.
To construct the perturbative target density operator entering the relative entropy, we normalize the corresponding truncation of $e^{-\beta H_g}$ by its trace and assume that the normalized truncation is positive up to terms beyond the retained order, so that relative-entropy non-negativity can be imposed consistently to the same order (see also Ref.~\cite{Conzinu:2026qgr}).
Nonanalytic instability contributions are absent at every finite order, while contributions not captured by the local weak-background expansion are assumed to remain parametrically subleading under assumption~(iii).
Beyond this regime, Borel--Laplace resummation alone need not give the complete result.
For a stable generic EFT, nonperturbative contributions not captured by the resummed perturbative sector may modify a negative result; such negativity therefore does not by itself imply a violation of the relative-entropy inequality for the complete theory. 
If the analytic continuation instead reaches an unstable background, the continued expression may no longer admit a density-operator interpretation. In one-loop QED, for example, its negative real part accompanies the independently known Schwinger instability.
The factorial-growth ansatz in assumption~(iii) is specific to the bounds derived here, whereas the relative-entropy construction is not restricted to factorial growth.
As shown in Ref.~\cite{Conzinu:2026qgr}, analogous methods apply to power-growing EFTs under the corresponding weak-background control assumptions.

\vspace{0.3\baselineskip}
\noindent{\bf Fermionic QED.} --- We now illustrate the general arguments above with fermionic QED.
The Lagrangian is
$\mathcal{L}=-\tfrac{1}{4}F_{\mu\nu}F^{\mu\nu}
+\bar{\psi}(i\gamma^\mu D_\mu-m)\psi$
with $D_\mu=\partial_\mu+i e A_\mu$.
The corresponding density operator is
$\rho_{\rm T}=\lim_{\beta\to\infty}e^{-\beta H_e}/Z_e$.
Since the interaction Hamiltonian is linear in the coupling $e$,
it can be written as
$V_e=e\,v^{(1)}$ with
$v^{(1)}=\int d^3x\,\psi^\dagger
(\gamma^0\gamma^iA_i+A_0)\psi$,
which implies
$(dW_e/de)_{e=0}={\rm Tr}\,[\rho_{\rm R}\,\beta\, v^{(1)}]$.
Symmetry requires this quantity to vanish, implying that ${\rm Tr}\,[\rho_{\rm R}\,\beta\, V_e]=0$ in Eq.~\eqref{eq:quanrel}.

In the present one-loop QED example, the effective action is first evaluated for real Euclidean gauge backgrounds and then analytically continued to the desired Minkowski background (see Sec.~\ref{supp:pert} of the Supplemental Material).
The continuation leaves the spatial magnetic components unchanged and the magnetic result real. 
For a real electric field, by contrast, the proper-time poles require the standard contour prescription and generate the Schwinger imaginary part.
This contribution is nonanalytic at $\hat{E}=0$ and absent at every finite perturbative order.
Within the finite-order weak-field treatment, we normalize the perturbative density-operator construction by its trace and take it to remain positive within perturbative accuracy.
Once the pole contribution is retained in the full one-loop expression, however, the analytically continued quantity becomes complex and no longer represents a relative entropy between positive, normalized states.
Because the proper-time representation captures the complete one-loop background dependence, the negative real part in this controlled example accompanies the independently established Schwinger instability and indicates that the analytically continued expression no longer admits the positive-density-operator interpretation required for relative-entropy non-negativity.

Detailed expressions for the effective action obtained via the proper-time method are given in Ref.~\cite{Conzinu:2026qgr}.
Applying this method to constant electric
($E^i=F^{i0}$) and magnetic
($B^i=-\epsilon^{ijk}F_{jk}/2$) background fields,
we obtain the effective action entering the relative entropy~\eqref{eq:quanrel}:
\begin{align}
W_e
= \int d^4x_{\rm E}
\left(
\epsilon_0
-\frac{1}{2}(\vec{E}^2-\vec{B}^2)
-\mathcal{L}_{\rm nonlin}
\right),
\label{eq:W_nonlin_ferm}
\end{align}
where $\epsilon_0$ denotes a background-independent vacuum-energy term that drops out of the relative-entropy difference, and $\mathcal{L}_{\rm nonlin}$ represents the one-loop nonlinear EFT correction to Maxwell theory after wave-function renormalization.
In Minkowski spacetime,
the nonlinear EFT Lagrangian in background fields reads
$\mathcal{L}_{\rm EFT}
= -\frac{1}{4}F_{\mu\nu}F^{\mu\nu}
+\mathcal{L}_{\rm nonlin}$.
In what follows, we focus on two representative cases:
a magnetic background
($\vec{E}=0$, $\vec{B}\neq0$)
and an electric background
($\vec{E}\neq0$, $\vec{B}=0$).

In particular, the nonlinear effect of the magnetic background is $
\mathcal{L}_{\rm nonlin}((e \hat{B})^2)
=\tfrac{m^4}{8\pi^2}\int_0^\infty
e^{-t}t^{-3}\,\mathcal{K}(e\hat B t)\,dt$, with
$\mathcal{K}(x)=-ix\cot(ix)+1-\tfrac{(ix)^2}{3}$, where the dimensionless operator is
$\hat B=|\vec B|/m^2$.
Symmetries imply that this nonlinear EFT contribution
is expressed as a power series in $(e\hat{B})^2$
({\it i.e.}, $o=e\hat{B}$ and $k=2$ in Eq.~\eqref{eq:qed_rel}).
Expanding the operator and truncating the series yields the perturbative result
$\mathcal{L}_{\rm nonlin}((e\hat{B})^2)\simeq
m^4\sum_{n=2}^N c_n (e\hat{B})^{2n}$, with $c_n=(4\pi^2)^{-1}\zeta(2n)
(i\pi)^{-2n}(2n-3)!$,
where $\lim_{n\to\infty}\zeta(2n)=1$.
This matches the nearest-pole large-order form underlying Eq.~\eqref{eq:resu_rel}
with $M=m$,
$C=1/4\pi^2$, $g=e$,
$o=e\hat{B}$,
$k=2$,
$n_{\rm min}=2$,
$\ell=3$, and $\tau_{\rm pole}=i\pi$.

We now compare the general result~\eqref{eq:resu_rel} with the one-loop proper-time result retaining the full dependence on the constant background:
\begin{widetext}
{\fontsize{9.6}{9.} \selectfont
    \begin{align}
      S(\rho_{\rm R}\|\rho_{\rm T})
    =
    \int d^4x_{\rm E}\, \mathcal{L}_{\rm nonlin}= \frac{m^4}{4\pi^2}\,\sum_{p=1}^\infty\,  \int d^4x_{\rm E}
     \begin{cases}
    \int_0^\infty e^{-t}t^{-3} \frac{(e\hat{B} t/p\,i\pi)^4}{1-(e\hat{B}t/p\,i\pi)^2}dt\,, \qquad \qquad \qquad \text{magnetic case} ~(e\hat{B}/i\pi)^2<0 \,,&
    \\
    \mathcal{P}\,\int_0^\infty e^{-t}t^{-3} \frac{(e\hat{E} t/p\,\pi)^4}{1-(e\hat{E}t/p\,\pi)^2}dt+i\,\frac{\pi}{2} \frac{e^{-\pi p/e\hat{E}}}{(\pi p/e\hat{E})^2}\,, \,\,  \text{electric case}~ (e\hat{E}/\pi)^2>0\,. &
     \end{cases}
     \label{eq:pos_M}
\end{align}}
\end{widetext}
The general result~\eqref{eq:resu_rel} is therefore consistent with
the nearest-pole contribution ($p=1$) in Eq.~\eqref{eq:pos_M}.
In the magnetic case,
the positivity condition~\eqref{eq:pos_mag},
$C(-1)^{n_{\rm min}}=1/4\pi^2>0$,
is satisfied.
This follows from the non-negativity of the relative entropy in the stable magnetic background.

The electric case follows from the analytic continuation
$\hat B \to i \hat E$
with $\hat E = |\vec E|/m^2$,
yielding $\tau_{\rm pole} = \pi$
in Eq.~\eqref{eq:resu_rel}.
At one loop, the proper-time result verifies the weak-background suppression condition in assumption~(iii), while the nearest-pole asymptotics $\zeta(2n)\to1$ gives $C = 1/4\pi^2 > 0$, consistently with the sign required by Eq.~\eqref{eq:pos_ele}.
Beyond this weak-background check, the same proper-time representation retains the full one-loop dependence on the constant background and permits a controlled comparison with Eq.~\eqref{eq:resu_rel}.
In the strong-electric-field regime, the negative principal-value real part occurs in a regime where the Schwinger imaginary part already establishes vacuum instability, and both originate from the same proper-time pole structure.
The negative real part can therefore serve as a model-specific diagnostic of nonperturbative physics.

\vspace{0.3\baselineskip}
\noindent{\bf Summary.} --- In this Letter, we study a class of nonlinear EFTs whose perturbative expansions exhibit factorial growth.
We focus on the subset of EFTs for which the relative entropy encodes an infinite tower of higher-dimensional operators, such as shift-symmetric scalar and pure-gauge theories, and derive bounds on nonlinear EFTs.
Within the adopted factorial-growth ansatz and weak-background assumptions, relative-entropy non-negativity constrains the sign of the assumed asymptotic coefficient growth.
Beyond the weak-background regime, a negative real part of the analytically continued resummed expression indicates a limitation of the resummed perturbative description and possible sensitivity to nonperturbative physics, including instabilities.
Constant-field QED at one loop provides a controlled example in which the proper-time result retains the full background-field dependence and the negative real part accompanies the Schwinger instability.
This framework provides a new information-theoretic route to constraining divergent EFT towers and probing their sensitivity to nonperturbative physics.

\begin{acknowledgments}
We thank the CERN Theory Department for financial support and for providing the environment in which this work was initiated.
DU is supported by grants from the ISF (No. 1002/23 and 597/24) and the BSF (No. 2021800).
PC acknowledges the support of INFN under the program ``QGSKY:Quantum Gravity in the SKY''.
\end{acknowledgments}

\twocolumngrid
\vspace{-8pt}
\section*{References}
\vspace{-10pt}
\def\bibsection{}
\bibliography{EH.bib}

\input{supplement}

\end{document}

%% file: supplement.tex
\clearpage
\onecolumngrid

\setcounter{equation}{0}
\setcounter{figure}{0}
\setcounter{table}{0}
\setcounter{section}{0}
\makeatletter
\renewcommand{\theequation}{S\arabic{equation}}
\renewcommand{\thefigure}{S\arabic{figure}}
\renewcommand{\thetable}{S\arabic{table}}
\renewcommand{\thesection}{S\arabic{section}}
\begin{center}
 \large{\bf 
Bounds on nonlinear effective field theories via resurgent relative entropy
 }\\
 Supplemental Material
\end{center}
\begin{center}
Pietro Conzinu, and 
Daiki Ueda
\end{center}

In this Supplemental Material, we provide details of results presented in Refs.~\cite{Cao:2022iqh,Cao:2022ajt,Ueda:2024cyf}, together with additional details of the analysis in the main text.
Sections~\ref{supp:1}–\ref{supp:3} review results from Refs.~\cite{Cao:2022iqh,Cao:2022ajt,Ueda:2024cyf}.
In Sec.~\ref{supp:1}, we derive Eq.~\eqref{eq:quanrel}, while Sec.~\ref{supp:2} gives an alternative expression for ${\rm Tr}\,\left[\rho_{\rm R}\, \beta\, V_g\right]$.
In Sec.~\ref{supp:pert}, we review the partition function evaluation and separate the auxiliary Euclidean evaluation, analytic continuation to Minkowski backgrounds, and the electric-background instability.
In Sec.~\ref{supp:3}, we relate the relative entropy to higher-dimensional operators in EFTs.
Sections~\ref{supp:Borel} and \ref{sec:app:pos} provide further details of the main analysis.

\section{Derivation of the Relative Entropy Formula}
\label{supp:1}
For completeness, we provide a derivation of Eq.~\eqref{eq:quanrel}; see Refs.~\cite{Cao:2022iqh,Cao:2022ajt,Ueda:2024cyf} for further details.
For two density operators $\rho_{\rm T}$ and $\rho_{\rm R}$, the relative entropy is defined as
\begin{align}
S(\rho_{\rm R}\|\rho_{\rm T})= {\rm Tr}\, \left[\rho_{\rm R}\,\ln\,\rho_{\rm R}-\rho_{\rm R}\,\ln\,\rho_{\rm T}\right].
\label{eq:rel_supp}
\end{align}
We take $\rho_{\rm T} = e^{-\beta H_g}/Z_g$ and $\rho_{\rm R} = e^{-\beta H_0}/Z_0$,
with $Z_g={\rm Tr}\,\left[e^{-\beta H_g}\right]$ and $Z_0={\rm Tr}\,\left[e^{-\beta H_0}\right]$.
Using $H_g = H_0 + V_g$, where $V_g=\sum_{n=1}^\infty g^n v^{(n)}$, we obtain
\begin{align}
S(\rho_{\rm R}\|\rho_{\rm T})
= W_0 - W_g + {\rm Tr}\,\left[
\rho_{\rm R}\, \beta\, V_g
\right],
\end{align}
where we used $\ln \rho_{\rm R}=-\beta H_0 -\ln Z_0$, $\ln \rho_{\rm T}=-\beta H_g -\ln Z_g$, and defined $W_0=-\ln Z_0$ and $W_g=-\ln Z_g$.
From $W_g=-\ln Z_g$, we obtain ${d W_g}/{dg}
=
{\rm Tr}\,\left[\rho_{\rm T}\,\beta \left({d V_g}/{dg} \right)\right]$.
Taking the limit $g\to 0$, we obtain
\begin{align}
\left(\frac{d W_g}{dg}\right)_{g=0}
=
{\rm Tr}\,\left[
\rho_{\rm R}\,\beta\, v^{(1)}
\right]\,,
\end{align}
where we used $\lim_{g\to 0}(dV_g/dg)=v^{(1)}$ and $\lim_{g\to 0}\rho_{\rm T}=\rho_{\rm R}$.

\section{Alternative Form of the Expectation Value}
\label{supp:2}
Following Refs.~\cite{Cao:2022iqh,Cao:2022ajt,Ueda:2024cyf}, we express the expectation value ${\rm Tr}\,\left[\rho_{\rm R}\, \beta\, V_g\right]$ in Eq.~\eqref{eq:quanrel} in an alternative form.
We introduce an auxiliary parameter $\lambda$ as
\begin{align}
H_\lambda \equiv H_0+\lambda\, V_g\,.
\label{eq:Hlambda}
\end{align}
Here, $\lambda$ is an auxiliary interpolation parameter rather than a physical coupling; we assume that the Hamiltonians and density operators along this path remain well defined and sufficiently regular for $0\leq\lambda\leq1$.
Defining $Z_\lambda\equiv{\rm Tr}\,\left[e^{-\beta H_\lambda}\right]$ and $W_\lambda\equiv-\ln Z_\lambda$, we obtain
\begin{align}
\frac{d W_\lambda}{d\lambda}
= \frac{1}{Z_\lambda}\,{\rm Tr}\,\left[e^{-\beta H_\lambda}\,\beta\,\left(\frac{d H_\lambda}{d\lambda}\right)\right]
= {\rm Tr}\,\left[\frac{e^{-\beta H_\lambda}}{Z_\lambda}\,\beta V_g\right]\,,
\end{align}
where we used $d H_\lambda/d\lambda=V_g$.
Evaluating at $\lambda=0$, we obtain
\begin{align}
\left(\frac{d W_\lambda}{d\lambda}\right)_{\lambda=0}
= {\rm Tr}\,\left[\rho_{\rm R}\,\beta\, V_g\right]\,,
\label{eq:exp_lambda}
\end{align}
where $\rho_{\rm R}=\lim_{\lambda\to 0} e^{-\beta H_\lambda}/Z_\lambda$.
Therefore, the expectation value ${\rm Tr}\,\left[\rho_{\rm R}\, \beta\, V_g\right]$ in Eq.~\eqref{eq:quanrel} corresponds to the first-order contribution of the interaction $V_g$ to the effective action $W_g=W_{\lambda=1}$.
Following Refs.~\cite{Cao:2022iqh,Cao:2022ajt,Ueda:2024cyf}, this illustrates that the relative entropy encodes an infinite tower of higher-dimensional operators in EFTs, such as shift-symmetric scalar theories and pure $SU(N)$ gauge theories.
For completeness, Sec.~\ref{supp:3} also provides further details.

\section{Partition Function Evaluation}
\label{supp:pert}
In this section, we assume that $\phi$ denotes a light field that admits a background configuration at low energies, with heavy degrees of freedom integrated out.
The zero-temperature partition function $Z_g$ is given by
\begin{align}
Z_g[\phi^{\rm cl}] = \lim_{\beta\to \infty}\, {\rm Tr}\,\left[e^{-\beta\, H_g[\phi^{\rm cl}]}\right],\label{eq:Zgscal}
\end{align}
where the trace is taken over the full Hilbert space and $\phi^{\rm cl}$ denotes a stationary background in Minkowski spacetime; for backgrounds reached through analytic continuation, $Z_g[\phi^{\rm cl}]$ below denotes the corresponding continued function.
The Euclidean quantities introduced below are auxiliary computational objects.
We evaluate $Z_g[\phi^{\rm cl}]$ in two steps.
First, we introduce its Euclidean counterpart $Z_g^{\rm E}[\phi^{\rm E,cl}]$ and compute it using the Euclidean path integral.
When this auxiliary Euclidean construction admits an interpretation in terms of positive, normalized density operators, the relative entropy constructed from the corresponding Euclidean quantities is non-negative.
Second, we obtain the effective-action and partition-function expressions for the desired Minkowski background by analytically continuing $\phi^{\rm E,cl}$ to $\phi^{\rm cl}$.
We describe these two steps in detail below.

\begin{itemize}
    \item {\it Euclidean path integral.} --- The partition function can be written in terms of the Euclidean light field $\phi^{\rm E}$ as
    \begin{align}
    Z_g^{\rm E}[\phi^{\rm E,cl}]
    =\int \mathcal{D}\phi^{\rm E}\, z_g^{\rm E}[\phi^{\rm E}],\qquad 
    z_g^{\rm E}[\phi^{\rm E}]
    \equiv
    \lim_{\beta\to \infty} {\rm Tr}_{\rm partial}\,\left[
    e^{-\beta\, H_g[\phi^{\rm E}]}
    \right],\label{eq:Eucl_path}
    \end{align}
    where the partial trace ${\rm Tr}_{\rm partial}$ is taken over the heavy degrees of freedom.  
    At leading order in the saddle-point approximation, the path integral is dominated by the stationary configuration $\phi^{\rm E,cl}$, yielding
    \begin{align}
    Z_g^{\rm E}[\phi^{\rm E,cl}] \simeq z_g^{\rm E}[\phi^{\rm E,cl}],
    \label{eq:Z_z}
    \end{align}
    where $\phi^{\rm E,cl}$ generally encodes the wave-function renormalization of the light field $\phi$.

    \item {\it Analytic continuation.} --- We analytically continue the Euclidean field to Minkowski spacetime to obtain the corresponding continued effective-action and partition-function expressions:
    \begin{align}
    W_g[\phi^{\rm cl}]
    = W_g^{\rm E}[\phi^{\rm E,cl}]\big|_{\phi^{\rm E,cl}\to\phi^{\rm E,cl}(\phi^{\rm cl})},
    \qquad
    Z_g[\phi^{\rm cl}]
    = Z_g^{\rm E}[\phi^{\rm E,cl}]\big|_{\phi^{\rm E,cl}\to\phi^{\rm E,cl}(\phi^{\rm cl})},\label{eq:S10}
    \end{align}
    where $W_g^{\rm E}=-\ln Z_g^{\rm E}$.  
    From Eq.~\eqref{eq:Z_z}, we obtain
    \begin{align}
    W_g[\phi^{\rm cl}]
    \simeq w_g[{\phi}^{\rm cl}]
    = -\ln z_g[{\phi}^{\rm cl}],\label{eq:S11}
    \end{align}
    where $w_g[{\phi}^{\rm cl}]\equiv-\ln z_g[{\phi}^{\rm cl}]$ and $z_g[{\phi}^{\rm cl}]\equiv z_g^{\rm E}[{\phi}^{\rm E,cl}]\big|_{{\phi}^{\rm E,cl}\to{\phi}^{\rm E,cl}({\phi}^{\rm cl})}$.
\end{itemize}

Equations~\eqref{eq:S10} and \eqref{eq:S11} analytically continue the Euclidean effective-action functions to the prescribed Minkowski background.
This continuation does not by itself guarantee that the resulting quantity retains an interpretation in terms of positive, normalized density operators.
In one-loop QED, the continuation to a static magnetic background remains on a real, stable branch, whereas continuation to a constant electric field encounters the proper-time poles and produces the Schwinger imaginary part.
This contribution is nonanalytic at $\hat{E}=0$, exponentially suppressed at weak field, and absent at every finite order of the local weak-field expansion.
The continued electric-background expression is therefore generally complex; a negative principal-value real part in the strong-field regime accompanies the independently known Schwinger instability rather than representing a negative relative entropy between positive, normalized states.

\section{Relative entropy as an infinite tower of higher-dimensional operators}
\label{supp:3}
We assume that nonlinear EFT effects arise from interactions between heavy and light fields.
Within this setup, we show that in EFTs such as shift-symmetric scalar theories and pure $SU(N)$ gauge theories, the relative entropy defined on suitable background fields takes the form of an infinite tower of higher-dimensional operators.

For later convenience, we introduce notation for the interaction terms (see also Refs.~\cite{Cao:2022iqh,Cao:2022ajt}).
For a light field $\phi$ and a heavy field $\Phi$, consider a UV theory with Lagrangian $\mathcal{L}=\mathcal{L}_0+\mathcal{L}_{\rm I}$, where the interaction term is written as
\begin{align}
\mathcal{L}_{\rm I}(\phi,\Phi)= \mathcal{O}(\Phi)\otimes J(\phi)\,,
\end{align}
where $\mathcal{O}(\Phi)\otimes J(\phi)$ involves implicit contractions of indices ({\it e.g.}, Lorentz indices).
In general, the operator $J(\phi)$ can be expanded as
\begin{align}
J(\phi)=J_{{\rm dim}\leq 4}(\phi)+\sum_{i=5,\ldots}\frac{1}{\Lambda^{i-4}}\, J_{{\rm dim}\text{-}i}(\phi)\,,\label{eq:J_def}
\end{align}
where $J_{{\rm dim}\leq 4}$ denotes operators of dimension up to four and $J_{{\rm dim}\text{-}i}$ denotes operators of dimension $i$.
Here $\Lambda$ is a mass scale satisfying $M\ll \Lambda$, with $M$ the mass of $\Phi$.
In what follows, we neglect $J_{{\rm dim}\text{-}i}(\phi)$ with $i>4$, as they are irrelevant for our purposes.
Allowing such terms would, in a generic Wilsonian setup, also permit independent higher-dimensional operators constructed solely from the light fields in $\mathcal{L}_0$, generated by physics at or above $\Lambda$.
To isolate the EFT effects induced by the restricted interaction
$\mathcal{L}_I=\mathcal{O}(\Phi)\otimes J_{\dim\leq4}(\phi)$, we consistently omit these additional higher-dimensional sectors.
Under this setup, we consider two representative EFTs: shift-symmetric scalar theories and pure $SU(N)$ gauge theories.

\subsection{Shift-symmetric scalar theory}
\label{sec:shift}
While the following discussion generalizes to multiple scalar fields (see Ref.~\cite{Ueda:2024cyf} for tree-level UV completions), we focus for simplicity on a single massless scalar field with a shift symmetry $\phi \to \phi + {\rm const.}$:
\begin{align}
\mathcal{L}_{\rm EFT}=\frac{1}{2}(\partial_\mu \phi)^2 +\mathcal{L}_{\rm nonlin}(\partial_\mu\phi)\,,
\end{align}
where $\mathcal{L}_{\rm nonlin}(\partial_\mu\phi)$ represents nonlinear EFT corrections depending on $\partial_\mu\phi$.
In this model, the allowed interaction terms in the UV theory (see also below Eq.~\eqref{eq:J_def}) are given by
\begin{align}
\mathcal{L}_{\rm I}\left(\phi,\Phi\right)= \mathcal{O}_\mu^{(1)}\left(\Phi\right) \partial^\mu\phi+ \mathcal{O}_{\mu\nu}^{(2)}\left(\Phi\right)(\partial^\mu \phi)(\partial^\nu \phi)\,.
\end{align}
Correspondingly, the UV Lagrangian is given by $\mathcal{L}_\lambda = \mathcal{L}_0 +\lambda\, \mathcal{L}_{\rm I}$, in analogy with the Hamiltonian~\eqref{eq:Hlambda}.

We consider the expectation value~\eqref{eq:exp_lambda} appearing in the relative entropy~\eqref{eq:quanrel}.
As explained in Sec.~\ref{supp:pert}, we evaluate the relative entropy in two steps: first in Euclidean spacetime using the path integral, and then via analytic continuation.
Specifically, we compute the relative entropy, equivalently the partition functions, in Euclidean spacetime and subsequently analytically continue the background light fields to Minkowski spacetime.
For concreteness, we evaluate the expectation value~\eqref{eq:exp_lambda} on a constant background field satisfying $\partial_\mu \phi=\mathrm{const.}$

Following the procedure in Secs.~\ref{supp:2} and~\ref{supp:pert}, in Euclidean spacetime we obtain
\begin{align}
    \left(\frac{dw^{\rm E}_\lambda[\phi]}{d\lambda}\right)_{\lambda=0} =\int \mathcal{D}\Phi^{\rm E} \int d^4x_{\rm E}\, \mathcal{L}^{\rm E}_{\rm I}\left(\phi,\Phi\right) \, \frac{e^{-\int d^4x_{\rm E} \mathcal{L}_0^{\rm E}}}{z_0^{\rm E}}=\int d^4x_{\rm E}\,\left(
    \langle\mathcal{O}_I^{(1)}\rangle_0\, \partial_I\phi+  \langle \mathcal{O}_{IJ}^{(2)}\rangle_0\,(\partial_I \phi)(\partial_J \phi)
    \right)\,,
\end{align}
where ${\langle \mathcal{O}\rangle}_0\equiv \int \mathcal{D}\Phi^{\rm E}\, \mathcal{O}\, e^{-\int d^4x_{\rm E}\mathcal{L}_0^{\rm E}}/z_0^{\rm E}$.
Here, by exploiting the $O(4)$ symmetry of Euclidean spacetime, we obtain
\begin{align}
    \langle\mathcal{O}_I^{(1)}\rangle_0=0,\qquad \langle\mathcal{O}_{IJ}^{(2)}\rangle_0\propto \delta_{IJ}.
\end{align}
By analytically continuing the background field, we obtain
\begin{align}
     \left(\frac{dw_\lambda[\phi]}{d\lambda}\right)_{\lambda=0} =-\frac{\alpha}{2}\, \int d^4x_{\rm E}\, (\partial_\mu\phi)^2\,,\label{eq:wlambda_scl}
\end{align}
where $\alpha$ denotes a constant.
Therefore, we obtain
\begin{align}
    w_\lambda[\phi]=\int d^4x_{\rm E}\left(-\frac{1}{2}\left(1+\lambda\,\alpha + \mathcal{O}\left(\lambda^2\right)\right)(\partial_\mu\phi)^2-\mathcal{L}_{\rm nonlin}\left(\partial_\mu\phi\right)\right)\,,\label{eq:wlambda_scl}
\end{align}
where the nonlinear correction $\mathcal{L}_{\rm nonlin}\left(\partial_\mu\phi\right)\sim \mathcal{O}\left(\lambda^2\right)$.
Evaluating Eq.~\eqref{eq:wlambda_scl} on the background field $\phi^{\rm cl}$ ({\it i.e.}, performing the path integral over the light field), we obtain 
\begin{align}
    W_\lambda[\phi^{\rm cl}]= \int d^4x_{\rm E}
    \left(
    -\frac{1}{2}\left(1+\lambda\,\alpha+ \mathcal{O}\left(\lambda^2\right) \right)(\partial_\mu\phi^{\rm cl})^2-\mathcal{L}_{\rm nonlin}\left(\partial_\mu\phi^{\rm cl}\right)
    \right)
    =
    \int d^4x_{\rm E}
    \left(
    -\frac{1}{2}(\partial_\mu\bar{\phi}^{\rm cl})^2-\mathcal{L}_{\rm nonlin}\left(\partial_\mu\bar{\phi}^{\rm cl}\right)
    \right),\label{eq:scl_nonli}
\end{align}
where $\bar{\phi}^{\rm cl}$ is the renormalized stationary background field dynamically chosen by the path integral over $\phi$.
By canonically normalizing $\partial_\mu\phi^{\rm cl}=\left(1+\lambda\,\alpha+ \mathcal{O}\left(\lambda^2\right) \right)^{-1/2}\,\partial_\mu\bar{\phi}^{\rm cl}$, the kinetic term is independent of $\lambda$.
Consequently, we obtain
\begin{align}
    \left(\frac{dW_\lambda[\phi^{\rm cl}]}{d\lambda}\right)_{\lambda=0}
    =
    -\int d^4x_{\rm E}\, \left(\frac{d\mathcal{L}_{\rm nonlin}\left(\partial_\mu\bar{\phi}^{\rm cl}\right)}{d\lambda}\right)_{\lambda=0}=0,
\end{align}
where we have used $\mathcal{L}_{\rm nonlin}\left(\partial_\mu\phi\right)\sim \mathcal{O}\left(\lambda^2\right)$.
Consequently, combining Eq.~\eqref{eq:quanrel} with Eqs.~\eqref{eq:scl_nonli}, \eqref{eq:exp_lambda} and $W_g=W_{\lambda=1}[\phi^{\rm cl}]$, we find that the relative entropy takes the form of an infinite tower of higher-dimensional operators:
\begin{align}
    S(\rho_{\rm R}\|\rho_{\rm T}) = \int d^4x_{\rm E}\,
    \mathcal{L}_{\rm nonlin}\left(\partial_\mu\bar{\phi}^{\rm cl}\right)\,.
\end{align}

\subsection{Pure $SU(N)$ gauge theory}
\label{sec:SUN}
The EFT of a pure $SU(N)$ gauge theory is given by
\begin{align}
    \mathcal{L}_{\rm EFT}=-\frac{1}{4}F^a_{\mu\nu} F^{a,\mu\nu}
    +
     \mathcal{L}_{\rm nonlin}\left(F^a_{\mu\nu}\right)\,,
\end{align}
where $F^a_{\mu\nu}=\partial_\mu A^a_\nu -\partial_\nu A^a_\mu +g\, f^{abc} A^b_\mu A^c_\nu$ is the field strength of the gauge field $A^a_\mu$, $g$ is the $SU(N)$ gauge coupling, and $\mathcal{L}_{\rm nonlin}\left(F^a_{\mu\nu}\right)$ represents nonlinear EFT corrections depending on $F^a_{\mu\nu}$. 
Here, Greek letters denote Lorentz indices, while italic letters label $SU(N)$ color indices. The totally antisymmetric structure constants $f^{abc}$ are defined by $[T^a, T^b] = i f^{abc} T^c$, where $T^a$ are the generators of the $SU(N)$ Lie algebra.
In what follows, we first focus on the pure $U(1)$ gauge theory EFT and then generalize the results to $SU(N)$.

\begin{itemize}
    \item {\it Pure $U(1)$ gauge theory} --- For a $U(1)$ gauge theory, the interaction terms in the UV theory are typically
\begin{align}
    \mathcal{L}_{\rm I}\left(A_\mu,\Phi\right)&= 
    \mathcal{O}^{(1)}_\mu\left(\Phi\right) A^\mu + \mathcal{O}^{(2)}_{\mu\nu}\left(\Phi\right) A^\mu A^\nu
    +
    \mathcal{O}^{(5)}\left(\Phi\right)\, (F_{\mu\nu}F^{\mu\nu})+\mathcal{O}^{(6)}\left(\Phi\right)\, (F_{\mu\nu}\widetilde{F}^{\mu\nu})\,.\label{eq:int_U1}
\end{align}
Regarding the terms $A^\mu$ and $A^\mu A^\nu$ in Eq.~\eqref{eq:int_U1}, we assumed the minimal gauge interactions arising from the covariant derivatives in the bilinear kinetic terms of the heavy charged fields, while neglecting cubic and quartic interactions of $A^\mu$ generated by higher-dimensional operators.
That is, in this setup, the operators $\mathcal{O}^{(2,5,6)}$ are invariant only under gauge transformations acting on the heavy charged fields.
In gauge theories, there is an ambiguity in the choice of the heavy–light-decoupled reference theory due to gauge symmetry.
The most straightforward definition of the heavy–light-decoupled reference theory is obtained by setting $\mathcal{L}_{\rm I}(A_\mu,\Phi)=0$, {\it i.e.}, $\mathcal{L}_0(A_\mu,\Phi)$.
However, for $A_\mu=\partial_\mu\alpha$, the interactions in Eq.~\eqref{eq:int_U1} do not yield physical effects at low energies, where the heavy fields are integrated out, due to gauge symmetry.
That is, by a field redefinition of the heavy field $\Phi\to \Phi'=U(\alpha)\Phi$ associated with the gauge transformation $U(\alpha)$, the interaction term $\mathcal{L}_{\rm I}(\partial_\mu\alpha,\Phi)$ can be absorbed into the non-interacting term $\mathcal{L}_0(A_\mu,\Phi')$, {\it i.e.},
$\mathcal{L}_0(A_\mu,\Phi')=\mathcal{L}_0(A_\mu,\Phi) + \mathcal{L}_{\rm I}(\partial_\mu\alpha,\Phi)$.
Thus, the theory $\mathcal{L}_0(A_\mu,\Phi) + \mathcal{L}_{\rm I}(\partial_\mu\alpha,\Phi)$ can be regarded as a heavy–light-decoupled reference theory.

By introducing this generalized reference theory and following the notation in Sec.~\ref{supp:2}, we define
\begin{align}
    \bar{\mathcal{L}}_\lambda\equiv \mathcal{L}_0 + \mathcal{L}_{\rm I} \left(\partial_\mu\alpha,\Phi\right)
    +
    \lambda\, \left(\mathcal{L}_{\rm I} \left(A_\mu,\Phi\right)-\mathcal{L}_{\rm I} \left(\partial_\mu\alpha,\Phi\right)\right)\,.\label{eq:barL}
\end{align}
For $\lambda=1$, Eq.~\eqref{eq:barL} reduces to the interacting theory, while for $\lambda=0$, it defines the reference heavy–light-decoupled theory. 
The Euclidean path integrals are then given by
\begin{align}
    Z_\lambda^{\rm E}[A^{\rm E,cl}]=\int \mathcal{D}A^{\rm E}\, z_\lambda^{\rm E}[A^{\rm E}],\qquad 
    z_\lambda^{\rm E}[A^{\rm E}]
    =
    \int \mathcal{D}\Phi^{\rm E}\, e^{-\int d^4x_{\rm E}\bar{\mathcal{L}}_\lambda},
\end{align}
with $Z_g^{\rm E}[A^{\rm E,cl}]=Z_{\lambda=1}^{\rm E}[A^{\rm E,cl}]$ and $z_g^{\rm E}[A^{\rm E}]=z_{\lambda=1}^{\rm E}[A^{\rm E}]$.
The remaining steps in the evaluation of the relative entropy proceed as in the previous case.

Following the procedure in Sec.~\ref{supp:pert}, we evaluate the derivative of $w_\lambda^{\rm E}=-\ln\,z_{\lambda}^{\rm E}[A^{\rm E}]$ in Euclidean spacetime and obtain
\begin{align}
    \left(\frac{dw^{\rm E}_\lambda}{d\lambda}\right)_{\lambda=0} &=\int \mathcal{D}\Phi^{\rm E} \int d^4x_{\rm E}\, \left(\mathcal{L}_{\rm I}\left(A_I,\Phi\right)-\mathcal{L}_{\rm I}\left(\partial_I\alpha,\Phi\right)\right) \, \frac{e^{-\int d^4x_{\rm E} \bar{\mathcal{L}}_0^{\rm E}}}{z_{\lambda=0}^{\rm E}}\notag
    \\
    &=\int d^4x_{\rm E}\,\bigg(c^{(2)}\left(|A^I|^2-|\partial^I\alpha|^2\right)+
    c^{(5)}\,F_{\mu\nu}F^{\mu\nu}+c^{(6)}\,F_{\mu\nu}\widetilde{F}^{\mu\nu}
    \bigg)\,,\label{eq:U1_dw_lam0}
\end{align}
where we omit the superscript E for the gauge field $A^I$.
Using the $O(4)$ symmetry of Euclidean spacetime, in Eq.~\eqref{eq:U1_dw_lam0}, we have used the following relations:
\begin{align}
    &{\langle \mathcal{O}_I^{(1)}\rangle}_0=0,\quad  {\langle \mathcal{O}_{IJ}^{(2)}\rangle}_0=c^{(2)}\, \delta_{IJ},\quad{\langle \mathcal{O}^{(5)}\rangle}_0= c^{(5)},\quad {\langle \mathcal{O}^{(6)}\rangle}_0= c^{(6)}\,,
\end{align}
with ${\langle \mathcal{O}\rangle}_0= \int \mathcal{D}\Phi^{\rm E}\, \mathcal{O}\, e^{-\int d^4x_{\rm E}\bar{\mathcal{L}}_0^{\rm E}}/z_0^{\rm E}$.
In general, deriving ${\langle \mathcal{O}_I^{(1)}\rangle}_0=0$ requires an appropriate choice of the gauge parameter $\alpha$ satisfying $\partial_I(A_I-\partial_I\alpha)=0$, as given below.
When the bilinear operator $\mathcal{O}_I^{(1)}$ is not invariant under gauge transformations of the heavy charged fields alone, its variation may be proportional to $\partial_I\alpha$.
Such a term, however, vanishes upon integration by parts and an appropriate gauge choice; see Ref.~\cite{Conzinu:2026qgr} for an explicit demonstration in scalar QED.
In what follows, we show that, by choosing suitable background fields and the gauge parameter $\alpha$, the contributions from $c^{(2)}$ vanish in Eq.~\eqref{eq:U1_dw_lam0}.

We begin by considering a constant magnetic field configuration, $\vec{B}=(0,\,0,\,B)$, as a simple illustrative case.
The following discussion generalizes straightforwardly to configurations including electric fields.
A corresponding gauge field configuration is given by $A_x=-(B/2)\,y$, $A_y=(B/2)\,x$, and $A_z=A_0=0$.
From this, the field strengths are
$B_x=F_{yz}=0$, $B_y=F_{zx}=0$, $B_z=F_{xy}=B$, and $E_k=F_{0k}=0$.
We then obtain
\begin{align}
    |A_I|^2= \frac{B^2}{4}\,\left(x^2+y^2\right)=\frac{B^2}{4}\, r^2\,,\qquad r^2=x^2+y^2\,.
\end{align}
We now consider solutions to $|\partial_I\alpha|^2=|A_I|^2=({B^2}/{4})\,r^2$.
For example,
\begin{align}
    \alpha =-\frac{B}{2}xy
    \,,\label{eq:gauge}
\end{align}
satisfies this condition and yields
\begin{align}
    \partial_I(A_I-\partial_I\alpha)=0
    \,.
\end{align}
Thus, with the gauge choice in Eq.~\eqref{eq:gauge}, both the contribution proportional to $c^{(2)}$ and the expectation value $\langle \mathcal{O}_I^{(1)}\rangle_0$ vanish.

Now, the remaining contribution in Eq.~\eqref{eq:U1_dw_lam0} is
\begin{align}
    \left(\frac{dw^{\rm E}_\lambda}{d\lambda}\right)_{\lambda=0} = \int d^4x_{\rm E}\, c^{(5)}\, F_{\mu\nu}F^{\mu\nu}\,, \label{eq:U1_dw}
\end{align}
where we have neglected the $c^{(6)}$ term, which can be removed by an appropriate choice of background field, {\it e.g.}, a purely magnetic configuration.
In this case, analogous to Sec.~\ref{sec:shift}, we have
\begin{align}
    w_\lambda[A] = \int d^4x_{\rm E} \left[
        -\frac{1}{4}\left(1 - 4\lambda\, c^{(5)} + \mathcal{O}(\lambda^2)\right) F_{\mu\nu}F^{\mu\nu} 
        - \mathcal{L}_{\rm nonlin}\left(F_{\mu\nu}\right)
    \right]\,, \label{eq:u1_wlambda}
\end{align}
where, from Eq.~\eqref{eq:U1_dw}, the nonlinear correction satisfies $\mathcal{L}_{\rm nonlin}(F_{\mu\nu}) \sim \mathcal{O}(\lambda^2)$.
Substituting the background field $A_\mu^{\rm cl}$ into Eq.~\eqref{eq:u1_wlambda} ({\it i.e.}, performing the path integral over $A_\mu$), we obtain
\begin{align}
    W_\lambda[A^{\rm cl}] = \int d^4x_{\rm E} \left[
        -\frac{1}{4} \bar{F}_{\mu\nu}\bar{F}^{\mu\nu} 
        - \mathcal{L}_{\rm nonlin}\left(\bar{F}_{\mu\nu}\right)
    \right]\,, \label{eq:Acl_nonli}
\end{align}
where $\bar{F}_{\mu\nu}$ is the renormalized stationary background field dynamically determined by the path integral over $A_\mu$.
By canonically normalizing $F_{\mu\nu}^{\rm cl} = \left(1- 4\lambda\, c^{(5)} + \mathcal{O}(\lambda^2)\right)^{-1/2} \bar{F}_{\mu\nu}$, the kinetic term becomes independent of $\lambda$.
Consequently, we find
\begin{align}
    \left(\frac{dW_\lambda[A^{\rm cl}]}{d\lambda}\right)_{\lambda=0} 
    = - \int d^4x_{\rm E} \left(\frac{d\mathcal{L}_{\rm nonlin}(\bar{F}_{\mu\nu})}{d\lambda}\right)_{\lambda=0} = 0,
\end{align}
where we have used $\mathcal{L}_{\rm nonlin}(F_{\mu\nu}) \sim \mathcal{O}(\lambda^2)$.
Combining Eqs.~\eqref{eq:quanrel}, \eqref{eq:Acl_nonli}, and \eqref{eq:exp_lambda} with $W_g = W_{\lambda=1}[A^{\rm cl}]$, we conclude that the relative entropy takes the form of an infinite tower of higher-dimensional operators:
\begin{align}
    S(\rho_{\rm R} \| \rho_{\rm T}) = \int d^4x_{\rm E}\, \mathcal{L}_{\rm nonlin}(\bar{F}_{\mu\nu})\,.
\end{align}
Note that the relative entropy generally depends on the choice of background light fields, which should be selected according to the specific information one aims to extract.

    \item {\it Pure $SU(N)$ gauge theory} --- As studied in Refs.~\cite{Cao:2022iqh,Cao:2022ajt}, the non-Abelian commutator contribution can be eliminated by choosing a color-aligned background, $A_{\mu}^a=u^aA_\mu$, where $u^a$ is a real constant vector in color space and $A_\mu$ is the Abelian background considered above.
Since $f^{abc}u^bu^c=0$, its field strength is $F_{\mu\nu}^a
=
u^a(\partial_\mu A_\nu-\partial_\nu A_\mu)$, which can remain nonzero.
The corresponding reference background is the color-aligned pure-gauge configuration $u^a\partial_\mu\alpha$.
The pure-$U(1)$ construction therefore applies to this color-aligned $SU(N)$ sector.

\end{itemize}

\section{Borel transformation}
\label{supp:Borel}
We consider the formal EFT expansion of an operator $o$,
\begin{align}
    \mathcal{L}_{\rm nonlin}(o^k)=M^4\, \sum_{n=n_{\rm min}}^\infty c_n\, o^{kn}\,,\label{eq:Borel_L}
\end{align}
where $n_{\rm min} \geq 1$ and $k \geq 1$.
In this Letter, we focus on the coefficients $c_n$, which exhibit factorial growth at large $n$:
$c_n \sim C\, (\tau_{\rm pole})^{-kn} (kn-\ell)!$.
For Eq.~\eqref{eq:Borel_L}, we define the Borel transformation as
\begin{align}
    \mathcal{B}(\tau) \equiv \sum_{n=n_{\rm min}}^\infty  \frac{c_n}{(kn-\ell)!}\,\tau^{kn-\ell}\,.
\end{align}
From this, we obtain
\begin{align}
    \int_0^\infty e^{-t}\, \mathcal{B}(o t) dt
    &\to
    \sum_{n=n_{\rm min}}^\infty \frac{c_n\, o^{kn-\ell}}{(kn-\ell)!} 
    \int_0^\infty e^{-t}\, t^{kn-\ell} dt
    =
    \sum_{n=n_{\rm min}}^\infty \frac{c_n\, o^{kn-\ell}}{(kn-\ell)!}\, \Gamma(kn-\ell+1)
    =
    \sum_{n=n_{\rm min}}^\infty c_n\, o^{kn-\ell}\,.\label{eq:trans_B}
\end{align}
Comparing Eq.~\eqref{eq:Borel_L} with Eq.~\eqref{eq:trans_B}, we thus obtain the resummed nonlinear effect:
\begin{align}
    \mathcal{L}_{\rm nonlin}(o^k) = M^4\, \int_0^\infty e^{-t}\,o^\ell \, \mathcal{B}(o t) dt\,. 
\end{align}

\section{Positivity in the weak-background regime}
\label{sec:app:pos}
In what follows, we determine the sign of the principal-value kernel multiplying $C$ in Eq.~\eqref{eq:resu_rel}.
It is positive in the weak-background regime but can become negative in the strong-background regime.
From Eq.~\eqref{eq:resu_rel}, we then obtain
\begin{align}
    {\rm Re}\,S(\rho_{\rm R}||\rho_{\rm T})\simeq C\, M^4\,\left|\frac{\tau_{\rm pole}}{o}\right|^{1-\ell}\, \int d^4x_{\rm E}\,\mathcal{F}(\tau_{\rm pole}/o)~~{\rm for}~(o/\tau_{\rm pole})^k>0,\quad
    \mathcal{F}(a)\equiv\mathcal{P}\,\int_0^\infty e^{-|a|t}\,\frac{t^{kn_{\rm min}-\ell} }{1-t^k}dt.\label{eq:Re_S_F}
\end{align}
The principal value integral defined above can be written explicitly as
\begin{align}
\mathcal{F}(a)=\lim_{\epsilon\to 0^+}\left[\int_0^{1-\epsilon}e^{-|a|t}\, \frac{t^{kn_{\rm min}-\ell}}{1-t^k}dt+\int_{1+\epsilon}^\infty e^{-|a|t}\, \frac{t^{k n_{\rm min}-\ell}}{1-t^k}dt\right].\label{eq:S42}
\end{align}
Using $t\to1/t$ and
$1/(1+\epsilon)=1-\epsilon+\mathcal{O}(\epsilon^2)$, we obtain
\begin{align}
\int_{1+\epsilon}^{\infty}
e^{-|a|t}
\frac{t^{kn_{\min}-\ell}}{1-t^k}\,dt
=
-\int_{0}^{1-\epsilon}
e^{-|a|/t}
\frac{t^{-k(n_{\min}-1)+\ell-2}}{1-t^k}\,dt
+\mathcal{O}(\epsilon),\label{eq:S43}
\end{align}
where the $\mathcal{O}(\epsilon)$ term vanishes in the principal-value limit.\footnote{
Near $t=1$, the transformed integrand behaves as
$e^{-|a|}/[k(1-t)]+\mathcal{O}(1)$, since
$1-t^k=k(1-t)+\mathcal{O}((1-t)^2)$.
Hence, its integral between $1-\epsilon$ and $1/(1+\epsilon)$ is
$e^{-|a|}\ln(1+\epsilon)/k+\mathcal{O}(\epsilon^2)
=e^{-|a|}\epsilon/k+\mathcal{O}(\epsilon^2)$,
which vanishes in the principal-value limit.
}
Combining Eqs.~\eqref{eq:S42} and \eqref{eq:S43}, we obtain
\begin{align}
    \mathcal{F}(a)=\lim_{\epsilon\to 0^+} \int_0^{1-\epsilon} t^{-\left(k(n_{\rm min}-1)-\ell+2\right)}\, \frac{\mathcal{N}(t,a)}{1-t^k}dt,\qquad \mathcal{N}(t, a)\equiv t^{2\kappa}\, e^{-|a|t}-e^{-|a|/t},
\end{align}
where $\kappa\equiv k\left(n_{\rm min}-1/2\right)+1-\ell$.
For $\mathcal{N}(t, a)>0$, it is evident that $\mathcal{F}(a)>0$. 
As we now show, on the pole-containing branch, the sign of $\mathcal F$ is controlled by two parameters: $\kappa$ and $|a|=|\tau_{\rm pole}/o|$:
\begin{align}
    \mathcal{F}(a)=\begin{cases}
    \geq 0\,, &\kappa =0
    \\
    > 0\,, & \kappa<  0~{\rm and}~|a|\ll 1\quad(\text{strong-background})
    \\
    <0\,, & \kappa>0~{\rm and}~|a|\ll 1\quad(\text{strong-background})
    \\
    >0\,, & \kappa\neq0, \quad |a/\kappa|\geq1  \quad(\text{weak-background})
    \end{cases}
\end{align}

First, for $\kappa=0$, the sign of $\mathcal{F}$ is independent of $|a|$ and is non-negative, since $\mathcal{N}(t, a)= e^{-|a|t}-e^{-|a|/t}\geq 0$.
We now consider two typical regimes: the strong-background regime ($|a|=|\tau_{\rm pole}/o|\ll 1$) and the weak-background regime ($|a|=|\tau_{\rm pole}/o| \gg 1$).

\begin{itemize}
    \item {\it Strong-background regime} --- In the limit $|a|=|\tau_{\rm pole}/o|\ll 1$ (strong-background regime), $\mathcal{N}(t,0)$, and hence $\mathcal{F}$ can become negative only for $\kappa>0$.
    This behavior is shown in Fig.~\ref{fig:Skln} of the main text.

    \item {\it Weak-background regime} --- In the weak-background regime $|a/\kappa|\geq1$ for $\kappa\neq0$, we show that $\mathcal{N}(t, a)>0$ holds.
Consider the following function:
\begin{align}
     n(t,a)&\equiv\ln \left(t^{2\kappa}\, e^{-|a|t}\right)
    -
    \ln \left(e^{-|a|/t}\right)=
    \kappa\,
    \left(
    \frac{|a|}{\kappa}\left(\frac{1}{t}-t\right)+2\ln t  
    \right)\,.
\end{align}
If $\mathcal{N}(t, a)>0$, we find $n(t,a)>0$.
Here, consider
\begin{align}
    m_+(t)=\frac{1}{t}-t +2\ln t,\qquad m_-(t)=-\left(\frac{1}{t}-t\right) +2\ln t,
\end{align}
For $0< t \leq 1$, we find
\begin{align}
    \frac{dm_+(t)}{dt}=-\frac{1}{t^2}-1+\frac{2}{t}=-\frac{(t-1)^2}{t^2}\leq 0,\qquad \frac{dm_-(t)}{dt}=\frac{1}{t^2}+1+\frac{2}{t}=\frac{(t+1)^2}{t^2}> 0.
\end{align}
Thus, $m_+(t)$ is monotonically decreasing, whereas $m_-(t)$ is monotonically increasing.
Since $m_+(1)=0$, $m_+(t)\geq 0$ for $0< t \leq 1$.
Moreover, since $m_-(1)=0$, $m_-(t)\leq 0$ for $0< t\leq 1$. 
We thus find
\begin{align}
    \left(\frac{1}{t}-t\right)+2\ln t &=m_+(t)\geq 0~{\rm for}~0< t \leq 1,\qquad -\left(\frac{1}{t}-t\right)+2\ln t =m_-(t)\leq 0~{\rm for}~0< t \leq 1.
\end{align}
For $\kappa>0$, and the weak-background regime $\left|{a}/{\kappa}\right|=\left|{\tau_{\rm pole}}/{\kappa\,o}\right|\geq 1$, we get
\begin{align}
    \left|\frac{a}{\kappa}\right| \left(\frac{1}{t}-t\right)+2\ln t \geq  \left(\frac{1}{t}-t\right)+2\ln t =m_+(t)\geq 0~{\rm for}~0< t \leq 1.
\end{align}
Similarly, for $\kappa<0$, and the weak-background regime $\left|{a}/\kappa\right|=\left|\tau_{\rm pole}/\kappa\,o\right|\geq 1$, we get
\begin{align}
     -\left|\frac{a}{\kappa}\right| \left(\frac{1}{t}-t\right)+2\ln t\leq - \left(\frac{1}{t}-t\right)+ 2\ln t =m_-(t)\leq 0~{\rm for}~0< t \leq 1.
\end{align}
That is, in the weak-background regime ($\left|{a}/{\kappa}\right|\geq 1$), we have $n(t,a)\geq 0$.
Thus, $\mathcal{F}(a)>0$ in this regime, so that ${\rm Re}\, S(\rho_{\rm R}\|\rho_{\rm T})$ has the sign of $C$.
Imposing perturbative relative-entropy non-negativity therefore gives $C>0$, as in Eq.~\eqref{eq:pos_ele}.

\end{itemize}